\documentclass[preprint,1p,times]{elsarticle}
\usepackage[utf8]{inputenc}
\usepackage[english]{babel} 
\usepackage{microtype}
\usepackage{graphicx}
\usepackage{subfigure}
\usepackage{booktabs} 
\usepackage{tipa}
\usepackage{upgreek}
\usepackage{lipsum}
\usepackage{amsmath}

\usepackage{float}
\usepackage{bm}
\usepackage{mathalpha}
\usepackage{siunitx}

\usepackage{amsmath,amssymb}
\usepackage{dsfont}
\usepackage{enumerate}
\usepackage{graphicx}      
\usepackage{comment}
\usepackage{algorithmic}
\usepackage{amsthm}

\DeclareMathAlphabet{\mathpzc}{OT1}{pzc}{m}{it}

\usepackage{mathtools}
\usepackage[ruled,vlined]{algorithm2e}
\usepackage{xcolor}

\begin{document}

\begin{frontmatter}

\title{Integrating process design and control using reinforcement learning}

\author[CPSE]{Steven Sachio}
\author[MU]{Max Mowbray}
\author[CPSE,IMSE]{Maria Papathanasiou}
\author[CPSE,IMSE]{Ehecatl Antonio del Rio-Chanona\corref{cor1}}
 \ead{a.del-rio-chanona@imperial.ac.uk}
\author[UCL]{Panagiotis Petsagkourakis\corref{cor1}}
 \ead{p.petsagkourakis@ucl.ac.uk}
 \cortext[cor1]{Corresponding authors}

\address[CPSE]{Sargent centre for Process Systems Engineering, Department of Chemical Engineering, Imperial College London, UK}
\address[IMSE]{Institute for Molecular Science and Engineering, Department of Chemical Engineering, Imperial College London, UK}
\address[UCL]{Sargent centre for Process Systems Engineering, Department of Chemical Engineering, University College London, UK}
\address[MU]{Department of Chemical Engineering and Analytical Science, University of Manchester, UK}

\date{2020}

\begin{abstract}
    To create efficient-high performing processes, one must find an optimal design with its corresponding controller that ensures optimal operation in the presence of uncertainty. When comparing different process designs, for the comparison to be meaningful, each design must involve its optimal operation. Therefore, to optimize a process' design, one must address design and control simultaneously. For this, one can formulate a bilevel optimization problem, with the design as the outer problem in the form of a  mixed-integer nonlinear program (MINLP) and a stochastic optimal control as the inner problem. This is intractable by most approaches. In this paper we propose to compute the optimal control using reinforcement learning, and then embed this controller into the design problem. This allows to decouple the solution procedure, while having the same optimal result as if solving the bilevel problem. The approach is tested in two case studies and the performance of the controller is evaluated. The case studies indicate that the proposed approach outperforms current state-of-the-art simultaneous design and control strategies. This opens a new avenue to address simultaneous design and control of engineering systems.

\end{abstract}

\end{frontmatter}

\section{Introduction}
\label{introduction}
    To improve operability, flexibility, reliability and economics of a process, design and control must be considered simultaneously~\citep{burnak2019}. Operational time-varying decisions include short-term process regulatory decisions (process control) and long-term economical decisions (scheduling). Recent studies have focused on the complete integration of design, control and scheduling to guarantee operability and profitability of operation under all foreseeable conditions~\citep{caspari2020, tsay2020, rafiei2020}.
    
    The scheduling problem addresses the allocation of resources, time, process unit utilization, and also maintenance. On the other hand, process control maintains or tracks the desired operating conditions under constraints. Due to the difference in objective and time scales, integration of design, scheduling and control remains a challenge. On a positive note, integration between design and control/scheduling has seen good progress~\citep{pistikopoulos2016}. For example, \citet{dirza2019} investigated the Max-Plus-Linear (MPL) representation for the integration of scheduling and control problem and \citet{diangelakis2017} proposed a multi parametric model predictive control (mp MPC) approach for simultaneous design and control.
    
    In this work, we focus on the integration of design and control, however, a similar approach can be extended to address scheduling, design and control. On one hand, the design problem aims to find specification and operating conditions. It is common to formulate this as a mixed-integer nonlinear program (MINLP). The objective function of the design problem is usually based on an economic performance index which accounts various factors such as operating and capital costs \citep{daoutidis2018}. 
    
    The control problem on the other hand, commonly known as the optimal control problem (OCP), is usually formulated as a dynamic optimization (DO) problem. It answers the question of how to adjust the manipulated process variables in order to maintain operation at the optimal condition. Therefore, in order to fully evaluate a design we would need to ensure that a control strategy exist such that the design can be maintained at the optimal operating condition, and that this ensures maximal performance. Similarly, to evaluate a control strategy, we need to apply it to a design. 
    
    This means that to fully optimize a process plant we need to consider both the design and control problems simultaneously, meaning that it would be necessary to couple both optimization problems. One way to address this problem, as stated before, is by nesting the control problem inside the design problem and solving the bilevel optimization problem directly~\citep{narraway1994}. This approach becomes intractable for even medium sized systems. Research has focused on the offline solution of the control problem using multi-parametric model predictive control (mp-MPC)~\citep{diangelakis2017}. However, the mp-MPC method requires reduction (mostly linearization) of the high-fidelity model to solve the control problem which may result in loss of information.  In this work, we propose to use an optimal policy computed via reinforcement learning, which has a closed-form expression and can be embedded into the outer (design) mathematical program as a mixed-integer (linear) constraint.
    
    \subsection{Previous approaches and state-of-the-art}
        One of the first structured methodologies to address simultaneous design and control, was presented by~\citet{narraway1991} called the “back-off approach”. This method can handle the design of linear and mildly nonlinear processes, and it can be broken down into three essential steps: (i) do steady-state nonlinear process optimization, (ii) linearize the problem at the optimum point and, (iii) “back-off” from the optimal solution to ensure that the process is feasible under disturbances. \citet{narraway1994} used this and formed an MINLP problem using a PI controller to solve the double effect evaporator problem. \citet{bahri1995} also employed this approach on optimal control to handle process uncertainties. 
        
        In 1992, the International Federation of Automatic Control (IFAC) organized the first workshop on simultaneous design and control at Imperial College London, which paved the way for the emergence of many different approaches to simultaneous control and design. For example, ~\citet{luyben1992} formulated a multi- objective optimization algorithm based on cutting planes involving open-loop controllability measures and economics, and~\citet{shah1992} implemented the State-Task Network (STN) scheme taking scheduling into consideration for multipurpose bath plants. At the time, literature about PI controllers was already abundant and well-established, furthermore, PI controllers are relatively easy to integrate due to their explicit form and became the most popular option to combine design and control. However, PI controllers cannot easily deal with nonlinear MIMO (multiple input multiple output) systems, for which MPC and its variants (e.g. dynamic matrix control) have become the industrial benchmark. 
    
       The advantage of MPC, relative to PI controllers, is based in the use of an explicit model of the process and postulation of a constrained dynamic optimization problem. One of the first significant contributions to implement MPC for simultaneous design and control was done by~\citet{brengel1992}. In this formulation, the authors implemented a bilevel optimization problem, where the outer level is the design problem formulated as an MINLP with an economic related objective function while the inner level is the control problem formulated as an MPC controller. Unfortunately, even with recent advances in bilevel programming \citep{djelassi2019, paulavicius2020}, the resultant bilevel optimization problem is very expensive to solve online and led to significantly fewer investigations utilising MPC to couple design and control, relative to the use of PI controllers.
       
       Since \citet{brengel1992}, several contributions have addressed the slow and expensive nature of schemes utilizing MPC to couple design and control. For example, \citet{chu2014} proposed a novel method utilizing a decomposition algorithm to solve bilevel integrated scheduling and optimization for sequential batch processes, while \citet{beykal2020} proposed a data-driven framework to tackle bilevel optimization problems. An approach to solve the bilevel problem is presented by \citet{bemporad2002, sakizlis2004a} where the authors replaced the inner problem with a multi-parametric model predictive controller (mp-MPC) reducing the online computation to a simple look-up table algorithm. In \citet{diangelakis2017} the authors worked further upon this approach and formulated a "design-dependent offline controller", where the implementation only requires solving the mp-MPC problem once offline. The biggest drawback of these mp-MPC approach is that they rely on linearizations, which do not well represent nonlinear dynamics, particularly when sampling times are long and multiple operational conditions are used. In another study by \citet{PALMAFLORES202011551}, the authors implemented nonlinear model predictive control (NMPC) for simultaneous design and control based on the back-off approach. Using NMPC, the authors managed to retain model accuracy at the cost of high computational demand. This can be feasible for small systems, however for large-scale systems the problem becomes intractable.
       
    \subsection{Contribution}
        In this work, we propose the use of reinforcement learning to solve the control problem offline and then couple it with the design problem. This can be done by taking advantage of the closed form (explicit) nature of reinforcement learning controllers. Furthermore, reinforcement learning (RL) controllers are a natural choice for dealing with chemical engineering problems, as they can address disturbances in highly nonlinear and complex processes without the need for linearizations~\citep{petsagkourakis2020}. Once the solution of a nonlinear optimal control problem is computed via RL (in the form of a feedback policy), it can be added as a constraint to the design problem, therefore avoiding the coupled inner minimization loop, while still having the same result as computing the solution simultaneously. Furthermore, if ReLU activation functions are used, this is a linear mixed-integer constraint. 
       
       In Section \ref{problemstatement}, an introduction to the formulation of the simultaneous design and control problem is presented. In Section \ref{preliminaries}, a brief overview of bilevel optimization, neural networks and policy gradients (reinforcement learning) is presented. Readers knowledgeable in these topics can proceed directly to Section \ref{methodandintegration}, where the methodology of the new approach is proposed. The results and discussion of the two case studies are presented in Section \ref{resultsanddiscussions}, Section \ref{conclusions} outlines the conclusion and future work.

%
%

\section{Problem Statement}
    \label{problemstatement}
        We formulate the simultaneous process design and control problem, as a bilevel program as shown in (\ref{eq0}) where the outer level is the design problem and the inner level is the control problem.
        \begin{equation}\label{eq0}
            \begin{split}
                \min_{\bm{p},\pi}~& C_p(\bm{p},\bm{x}(\tau), \bm{u}(\tau), \theta^{P}(\tau),\pi)\\
                s.t.~& h_{p}(\bm{p}, \theta^P(\tau),\pi)=0\\
                & g_{p}(\bm{p}, \theta^P(\tau),\pi)\le0\\
                &\begin{split}
                    \min_{\pi}~& C_{u}(\bm{x}(\tau), \bm{u}(\tau), \bm{p}, \theta^{P}(\tau))\\
                    s.t.~& \dot{\bm{x}}(\tau) = f(\bm{x}(\tau), \bm{u}(\tau), \bm{p}, \theta^{P}(\tau))\\
                    & \bm{u}(\tau) = \pi(\bm{x}(\tau), \bm{p})\\
                    & g_u(\bm{x}(\tau), \bm{u}(\tau), \bm{p}, \theta^{P}(\tau)) = 0\\
                    & h_u(\bm{x}(\tau), \bm{u}(\tau), \bm{p}, \theta^{P}(\tau)) \le 0
                \end{split}
            \end{split}
        \end{equation}
        where \(\tau\) is continuous time, \(\bm{x} \in \mathbb{R}^{n_{\bm{x}}}\) is the vector of states of the system, \(\bm{u} \in \mathbb{R}^{n_{\bm{u}}}\) is the control action described by the policy $\pi:\mathbb{R}^{n_x} \rightarrow \mathbb{R}^{n_{\bm{u}}}$, $\bm{p}:=\{\bm{z}, \bm{d}\}$ is the design variable vector including binary $\bm{z} \in \{0,1\}^{n_{\bm{z}}}$ and  continuous variables $\bm{d} \in \mathbb{R}^{n_{\bm{d}}}$, and \(\theta^P\) is the parameter vector of the process. \(C_{p}\) and \(C_{u}\) are the economic and control cost functions.
        Note that we assume to measure the full state \(\bm{x}\) for simplicity, and the control objective generally consists of an integral over time.

        The controller is represented by the policy $\pi$ parameterized by a set of parameters $\theta$ (e.g. $k_p,k_I$ for a PI controller). The policy could be given as a linear controller, e.g. \citep{chu2014,diangelakis2017}, however, this is suboptimal for most nonlinear dynamic systems, and therefore linear approximations are made. In this paper, we avoid the use of linear approximations in the dynamics and instead utilize a nonlinear artificial neural network (ANN) as a policy. To construct this neural-controller, reinforcement learning is used. Specifically, in this work, we propose to use policy gradients as the policy optimization approach to construction of an optimal policy $\pi_\theta$, where we use $\theta \in \mathbb{R}^{n_\theta}$ to denote the weights and bias of the neural network.

\subsection{The optimal control problem (OCP)}

        To construct a feedback control law in the form of a neural network controller, an optimal control problem (OCP) is formulated. The discrete-time formulation is presented in (\ref{eq2}).
        
        \begin{equation}\label{eq2}
            \begin{split}
                \min_{\pi_{\theta}}~& C_{u}(\cdot):=\mathbb{E}[J_{OCP}(\bm{x}_{t},\bm{u}_t,\bm{p})]\\
                s.t.~& \bm{x}_0 = \bm{x}(0) \\
                & \bm{x}_{t+1} = f(\bm{x}_{t}, \bm{u}_{t}, \bm{p}, \bm{w}_t)~\forall t\\
                & \bm{u}_t = \pi_{\theta} (\bm{x}_{t},\bm{x}_{t-1},\bm{p})\\
                & [\bm{x}_{min}, \bm{u}_{min}] \le [\bm{x}_t,\bm{u}_t] \le [\bm{x}_{max}, \bm{u}_{max}]~\forall t\\
                & t \in \{ 1,..., n_T-1 \}\\
            \end{split}
        \end{equation}
        
        where \(J_{OCP}\) is the cost function of the OCP (e.g. setpoint error)  and $\bm{w}_t\in \mathbb{R}^{n_{\bm{w}}}$ are random disturbances. The goal of this optimal control problem is to find the parameter values of the policy such that it optimizes the objective function. To solve this problem the policy is parameterized by an ANN (denoted, $\pi_\theta$) and then a policy gradient method is used \citep{NIPS1999_464d828b,petsagkourakis2020} to solve the optimal control problem.
         
        The overall problem is therefore a bilevel program, with the outer level being an MINLP and the inner level an OCP. By using RL, the bilevel problem can be separated and formulated as presented in (\ref{eq2}). Notice that the OCP takes into account the design variables to calculate the control output. When this is solved by RL, the final form of the policy is an explicit function, that takes states and design variables as input, and outputs an optimal control action, this can be embedded into the MIDO problem formulated in (\ref{eq0}).
        \emph{Remark: If ReLu activation functions are used, the resulting neural policy ($\pi_{\theta}$) result in a mixed integer linear function, which can be easily appended to the MINLP.}
        
        It is important to highlight that we are making approximations to solve the OCP (RL is an approximate dynamic programming technique). Furthermore, there is no guarantee that the solution is global, which might be important in some bilevel programming instances \citep{Mitsos2009}, however, guaranteeing global optimality in dynamic optimization is itself intractable in most problem instances \citep{FLORESTLACUAHUAC20061227}. 






\section{Preliminaries}
    \label{preliminaries}
    In this section we outline concepts that are relevant for this research. Extensive detail of the topics is not provided. For an extensive description of the topics the reader is directed to relevant literature subsequently.
    
    \subsection{Bilevel optimization}
        A bilevel optimization problem is an optimization problem that has another problem nested within as a constraint \citep{paulavicius2020}. Let us consider a simple bilevel problem shown in (\ref{eqbi}).
    \begin{equation}\label{eqbi}
            \begin{split}
                \min_{x,u}~& f_{1}(x, u)\\
                s.t.~& h_{1}(x,u) = 0\\
                & g_{1}(x,u)\le0\\
                &\begin{split}
                    \min_{u}~& f_{2}(x, u)\\
                    s.t.~& h_{2}(x, u) = 0\\
                    & g_{2}(x, u)\le0\\
                \end{split}
            \end{split}
        \end{equation}
        
        In the design and control case, \(x\) is the design variable, \(u\) is control variable, \(f_1\) is the design objective (e.g. operating cost), \(h_{1}\) and \(g_{1}\) are the equality and inequality constraints of the design problem respectively (e.g. maximum column length), \(f_2\) is the control objective (e.g. error for setpoint tracking), and \(h_{2}\) and \(g_{2}\) are the equality and inequality constraints of the control problem respectively (e.g. maximum temperature).

        The outer level problem is the design problem whose goal is to optimize an economic related objective function, while the inner level problem is the control problem whose goal is to find the optimum control actions to maintain optimal operating condition given the design variables \citep{diangelakis2017}. To solve the bilevel problem directly, we would need to explore different values of design variables and at each different design we need to solve the control problem. For large problems with high number of variable and constraints, direct solution of the bilevel problem quickly becomes intractable \citep{djelassi2019}.
        
        In this work, the problem tackled is similar to (\ref{eqbi}) but with more variables, variable types, equations and including the time dimension. Using traditional approach to solve this problem is intractable, there in this work we propose to take advantage of the closed form (explicit) nature of reinforcement learning controllers. For more details on bilevel optimization, the reader is directed to \citet{paulavicius2020}.
        
        
    \subsection{Reinforcement Learning}
        Reinforcement learning (RL) is a technique to optimize Markov Decision Process (MDP), given some performance metric. MDPs describe sequential decision-making problems in discrete-time stochastic processes, which posses the Markov property. This makes MDPs a good representation of stochastic dynamic systems, such as those in control engineering. Reinforcement learning is a method to address these systems \citep{petsagkourakis2020}. Consider the time discretization of a continuous-time problem, where \(\tau\) is the continuous time \(\tau\in[0, T_F]\), \(T_F\) is the time horizon which is discretized into \(n_T\) time steps of size \(\Delta\tau=T_F/n_T\). The subscript \(t\) denotes the time-step (i.e. \(\tau=t \Delta\tau\)), where \(t\in\{0,~1,...,~n_T-1,~n_T\}\). 
        
        In the beginning, the environment (or process) is at some initial state, \(\bm{x}_{t}\). The agent (or controller) is able to observe the state of the environment, \(\bm{x}_{t}\). The agent (or controller) then computes the policy, \(\pi_{\theta}\), which tells the agent which action, \(\bm{u}_t=\pi_{\theta}(\bm{x}_t)\), to take depending on the current state, \(\bm{x}_{t}\), of the environment. After the action has been chosen, the environment shifts to another state based on the transition probability function, \(p(\bm{x}_{t+1}|\bm{x}_t,\bm{u}_t)\). This probability function defines the dynamics of the process \citep{sutton2018}. Note that during training, the policy can be stochastic (e.g. \(\bm{u}_t \sim \pi_{\theta}(\bm{x}_t)=p(\bm{u}_t|\bm{x}_t) \))
        
\begin{figure}[hbt]
        \centering
        \includegraphics[width=9cm]{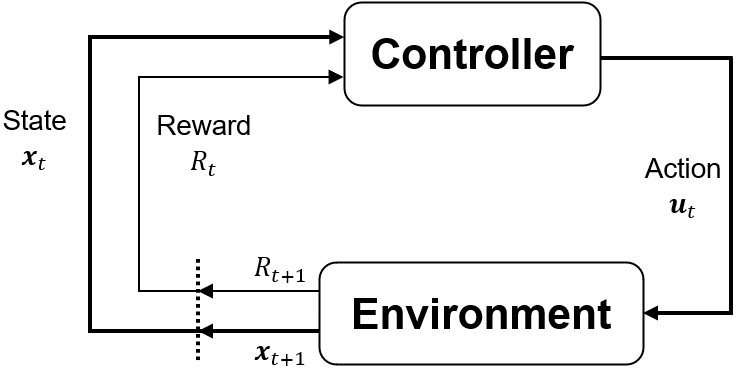}
        \caption{Interaction between the controller (agent) and the environment in an MDP.}
        \label{RL}
    \end{figure}

        In the context of RL, while training, the agent receives a 'reward', which is equivalent to a signal given to a controller which tells it about its performance. This reward is generally assumed to be produced by a reward function and it is defined by the operator. It is usually a measure of performance which the agent will try to maximise. For example, in a set point control problem, we can define a reward function to be the negative sum of the mean squared error. This would mean that the maximum reward that the agent can get will be equal to 0. This process is then repeated until the end of the defined time horizon. Shown in Fig. \ref{RL} is a block diagram illustration of the interaction between the agent and the environment. 
        
         The goal of the agent is to find the policy which gives the maximum reward. Therefore, we would need a methodology to improve our policy as we train the agent. In this work, we use policy gradients to improve our policy. More details on this are presented in the next section.

        \subsection{Policy Gradient}
        Policy Gradients (PG) is a methodology to estimate gradients of the reward (objective function) with respect to the parameters (say of a neural network's parameters) by sampling a stochastic function or environment. In this way, following a similar procedure to Gradient Descent (or it's variants for example: stochastic gradient descent \citep{ruder2016overview} and Adam \citep{kingma2017}), the RL-agent can 'move' towards a better parameterized policy until it reaches an optimal policy. The policy, \(\pi_{\theta}\), calculates the control action, \(\bm{u}_t=\pi_{\theta}(\bm{x}_t)\), given the current state, \(\bm{x}_{t}\in R^{n_x}\), and parameters, \(\theta\). For a stochastic policy (which can help exploration), the output from the neural network would be the mean and standard deviation of the control action. Then the actual control would be sampled from a distribution with the output mean and std \citep{petsagkourakis2020}.
    \begin{equation}\label{eqRL04}
        \bm{u} \sim \pi_{\theta}(\bm{u}|\bm{x}) = \pi(\bm{u}|\bm{x},\theta) = p(\bm{u}_t = \bm{u}|\bm{x}_t = \bm{x}, \theta_t = \theta)
    \end{equation}
        
        In PGs, the network parameters are updated using a gradient ascent-type strategy. 
        
        Let us consider the most basic PG algorithm, Reinforce, where Monte Carlo runs are obtained to compute state, control and reward trajectories. In this work we denote \(\bm{\kappa}\) as the joint variable of state, control and rewards trajectories with a time horizon of \(T_F\).
    \begin{equation}\label{eqRL05}
        \bm{\kappa} = (\bm{x}_0, \bm{u}_0, \bm{x}_1, R_1, \dots, \bm{x}_{n_T - 1}, \bm{u}_{n_T - 1}, \bm{x}_{n_T}, R_{n_T})
    \end{equation}
    
    Then, the return from the initial state, \(J(\bm{\kappa})\) (objective criterion) and the probability density of a trajectory can be expressed as shown in \ref{eqRL06} and \ref{eqRL07} \citep{sutton2018, petsagkourakis2020}.
    \begin{equation}\label{eqRL06}
    J(\bm{\kappa}) = \sum^{n_T-1}_{t=0} \gamma^t R_{t+1} (\bm{x}_{t}, \bm{u}_{t}, \bm{x}_{t+1})
    \end{equation}
    \begin{equation}\label{eqRL07}
    p(\bm{\kappa}|\theta) = \hat{\mu}(\bm{x}_0) \prod^{n_T-1}_{t=0} [\pi (\bm{u}_t|\bm{x}_t, \theta)p(\bm{x}_{t+1}|\bm{x}_t,\bm{u}_t)]
    \end{equation}
    
    where \(\hat{\mu}(\bm{x}_0)\), is the probability density of the initial state. The traditional gradient update rule is shown in \ref{eqRL08} \citep{sutton2018, petsagkourakis2020}.
    \begin{equation}\label{eqRL08}
    \theta_{m+1} = \theta_m + \alpha_m \hat{J}(\theta)
    \end{equation}
    \begin{equation}\label{eqRL09}
    \hat{J}(\theta) = \nabla_{\theta} \mathbb{E}_{\bm{\kappa} \sim p(\bm{\kappa}|\theta)} [J(\bm{\kappa})]
    \end{equation}
    
    where \(\alpha\) is the learning rate/step-size, \(m\) is the current iteration at which the parameters are updated (epoch). However, computing the gradient  of the expectation, \ref{eqRL09}, directly is hard. Hence, the \emph{Policy Gradient Theorem} \citep{sutton2018} is used. Therefore, the final update rule for this Reinforce Monte Carlo algorithm is shown in \ref{eqRL13} \citep{petsagkourakis2020}.
    \begin{equation}\label{eqRL13}
    \theta_{m+1} = \theta_m + \alpha_m \frac{1}{K} \sum^{K}_{k=1} [J(\bm{\kappa}^k) \nabla_\theta \sum^{n_T-1}_{t=0} \log(\pi(\bm{u}_t^k|\hat{\bm{x}}_t^k,\theta))]
    \end{equation}
        
    where the superscript \(k\) corresponds to sample number \(k\) in \(K\) total number of samples. Past work on using PG method in process control has seen success in dealing with nonlinear highly complex processes \citep{petsagkourakis2020, yoo2021}. In both studies, the approach was compared to nonlinear MPC and both studies showed that PG-based RL controllers have similar performance. The PG method, compared to other value-based reinforcement learning techniques, has a big advantage in the form of the end result. In value based methods, such as Q-learning, the end results is a table/function of action-value pairs (more details in \citep{sutton2018}) which needs another optimization to find out the best control action given a state of the system. In PG, on the other hand, the end result is a function taking in states of the system and giving out the optimal control action without the need for another optimization step. This helps in solving the bilevel optimization problem as the control problem reduces to a function-lookup problem. For more details on reinforcement learning and the policy gradient method, the reader is directed to \citet{sutton2018}.

\section{Method and Integration}
\label{methodandintegration}
    \subsection{Controller Design via Reinforcement Learning}
    \label{constructingcontroller}
        The neural network policy is computed based on the OCP formulated in Problem (\ref{eq2}). As mentioned previously, in this work we used PGs to optimize the controller. The controller took the form of a feed-forward neural network. However, this could be exchanged for any neural network architecture (e.g. RNN, LSTM, Transformers), or any other function parametrization. To speed-up the RL process, a pre-training (supervised learning) scheme is used \citep{torabi2018behavioral,ALMM21}. The main idea behind pre-training is that labelled data (\(X, U\)) are fed into the algorithm and the algorithm tries to find a relationship between them and then predict the output (\(\mathbf{u}_i \in U\)) given the state (\(\mathbf{x}_i \in X\)). In this case, we can use some simple policy (e.g. PI controller) to obtain this preliminary policy. This also allows us to use any other pre-existing policy.
        
        This means that in a process control context, the data could be a trajectory of states (\(X_{PT}\)) and the control actions chosen at each state (\(U_{PT}\)). Then, the neural network is trained so that if it encounters a state close to \(\bm{x}_{i} \in  X_{PT}\), it would return a control action which is close to the control \(\bm{u}_{i} \in  U_{PT}\). The algorithm is shown in Algorithm~\ref{alg:alg1}. 
        
        \begin{algorithm}
        \caption{Pre-training (supervised learning).}\label{alg:alg1}
        \smallskip
        \textbf{Input}: Initialize policy \(\pi_\theta\), with parameters \(\theta = \theta_0\), with \(\theta_0\in\Theta\), learning rate \(\alpha\), number of iterations \(n_{iter}\) and demonstration policy \(\pi_{demo}\) (e.g. PID controller).\\
        \textbf{Output}: Pre-trained policy.
        \smallskip
        
        {\bf for} m = 1,\dots, \(n_{iter}\) {\bf do:}
        \begin{enumerate}
                    \item Generate demonstration state trajectory \(X_{demo}\) and control actions \(U_{demo}\) using the demonstration policy.
                    \item Generate agent control actions \(U_{agent}\) using the current agent policy \(\pi_\theta\) based on the demonstration states \(X_{demo}\).
                    \item Improve \(\pi_\theta\) using \((U_{agent}, U_{demo})\) \citep{kingma2017}.
                    \item \(m:= m + 1\)
        \end{enumerate}
        \end{algorithm}
    
    After pre-training, the controller is deployed into the environment to begin the reinforcement learning. The Reinforce \citep{williams1992} (Monte Carlo) algorithm was used with the addition of baseline and decaying learning rate, although more complex schemes can be implemented \citep{petsagkourakis2020chance, schulman2017trust}. The algorithm is shown in Algorithm~\ref{alg:alg2}.
        
        \begin{algorithm}[H]

        \caption{Policy Gradient Algorithm.}\label{alg:alg2}
        \smallskip
        
        {\bf Input:} Initialize policy parameter $\theta = \theta_0$, with $\theta_0\in\Theta$, 
        learning rate $\alpha$, $m:=0$, the number of episodes $K$ and the number of epochs $N$, stochastic environment or process.\\
        {\bf Output:} policy $\pi(\cdot | \cdot ,\theta)$ and $\Theta$
        \smallskip
        
        {\bf for} m = 1,\dots, N {\bf do:}
        \begin{enumerate}
        \item Apply policy $\pi(\cdot | \cdot ,\theta_m)$ to process. Collect $\textbf{u}_t^k , \textbf{x}_t^k$ for $T$ time steps for $K$ trajectories along with $J(\textbf{x}_T^k)$, also for $K$ trajectories.
        \smallskip
        \item Update the policy network: $\theta_{m+1} = \theta_m + \alpha_m \frac{1}{K} \sum_{k=1}^{K}  \left[ (J(\pmb{\kappa}^k)-b) \nabla_\theta \sum_{t=0}^{T-1} \text{log}\left( \pi(\textbf{u}_t^k|\hat{\textbf{x}}_t^k,\theta)\right)\right]$
        \smallskip
        \item  $m:=m+1$
        \end{enumerate}
        \end{algorithm}

    The update rule used in the algorithm (step 2 of Algorithm \ref{alg:alg2}) is derived from the policy gradient theorem~\citep{sutton2018}. In this update rule, \(b\), is the baseline. Due to the stochasticity of the policy coupled with Monte Carlo sampling, the approximation of the expectation can have high variance, however, a baseline can be used to reduce this variance without a bias \citep{sutton2018}. The baseline used in this work is the expectation of rewards under the current policy, this is the most commonly used baseline in policy gradient projects and it has been proven to be effective \citep{petsagkourakis2020, yoo2021}.
    
    While the baseline helps to reduce the variance, the decaying learning rate on the other hand helps the training to be faster during the early epochs, and more stable during the later epochs. In this work we use Reinforce for simplicity of presentation, however, other policy gradient-based algorithms (e.g. TRPO \citep{schulman2017trust}, PPO \citep{schulman2017proximal}) can be used.
    
    \subsection{Integration}
    \label{integration}
    
    \begin{figure}
        \centering
        \includegraphics[scale=0.75]{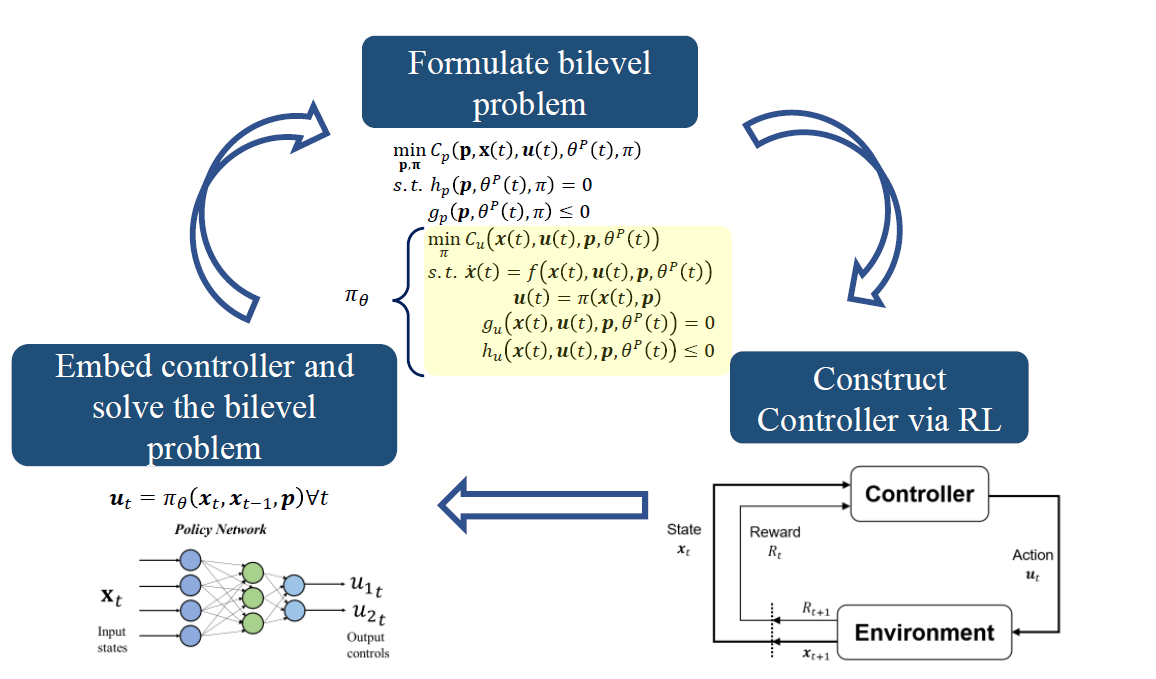}
        \caption{Simultaneous Design and Control Optimization using RL methodology.}
        \label{flowsheet}
    \end{figure}

    To summarize the methodology, a schematic representation of the proposed approach is presented in Figure \ref{flowsheet}. A bilevel optimization problem is formulated and split into two problems, the design problem and the control problem. The control problem is solved via reinforcement leaning, such that an optimal policy is produced. This optimal policy is a controller in the form of an ANN such that it takes states as inputs and outputs optimal control actions. Second, the policy is embedded into the design problem and solved using standard mixed-interger programming (MIP) methods. It is important to highlight, that we need a policy based RL method, if we were to use a value-based RL method, the OCP is an optimization problem, hence the integrated formulation in (\ref{eq0}) is again an intractable bilevel optimization problem. 
    
        %
        \begin{algorithm}
        \caption{Full approach for simultaneous design and control optimization using RL.} \label{alg:alg3}

        \textbf{Input}: Simultaneous process design and control problem.\\
        \textbf{Output}: Optimal design and controller.\\
         \begin{enumerate}
             \item {\bf Formulate simultaneous design-control optimization}: formulate the design problem as a mixed-integer dynamic optimization (MIDO) problem as presented in (\ref{eq0}).
            \smallskip
            \item {\bf Optimal Control problem}: Pose the optimal control problem (OCP) based on the design problem as in (\ref{eq2}).
            \smallskip
            \item \textbf{ OCP as RL}: Solve the OCP via RL.
            \smallskip
            \begin{enumerate}
                \item \textbf{Pre-training}: pre-train from an initial policy either via simulation or plant data, this can be done via supervised or apprenticeship learning, amongst others. The pre-training algorithm used in this study is shown in Algorithm \ref{alg:alg1}.
                \item \textbf{Policy learning via policy gradients}: Begin full- training using a policy optimization algorithm. The policy learning algorithm used in this study is shown in Algorithm \ref{alg:alg2}.
            \end{enumerate}
            \item \textbf{Embedded bilevel program}: Embed the final policy obtained from \textbf{step 3} into the design (MIDO) problem and solve via a MIP algorithm. 
         \end{enumerate} 
         \end{algorithm}

\section{Results and Discussions}
\label{resultsanddiscussions}
    \subsection{Tank Design}
    \label{tankdesign}
         In the first case study, we design a simple tank with continuous inlet and outlet flows. There are no reactions involved, the inlet flow is in the form of a sinusoidal signal and it has a nominal inlet flow of \(\hat{F}\) with a maximum deviation of \(F_d\). The outlet flow is the manipulated variable of the RL controller. A set-point which depends on the values of \(\hat{F}\) and \(F_d\) is taken as input to the controller. The design problem is formulated as a MIDO problem presented in (\ref{eq3}). In order to reduce the size of the problem while still maintaining a sampling time of 0.01 s to capture the dynamics of the sinusoidal signal, the final time is set to allow only one period of the sinusoidal signal.
    \begin{equation}\label{eq3}
        \begin{split}
            \max_{\overline{V}, F_d, \hat{F}, V(0)}~ &J_{SDC} = \int_0^{1}F_d~d\tau\\
            \text{subject to:}\\
            &\dot{V}(\tau) = F_{in}(\tau) - F_{out}(\tau)\\
            & F_{out}(\tau) = \alpha_t   V(\tau)\\
            & F_{in}(\tau) = \hat{F} + F_d   \sin{(2\pi \tau)}\\
            & V_{SP} = \hat{F} + F_d \le \overline{V}\\
            & \varepsilon_{\pi_\theta} = \int_0^1 \frac{||V(\tau)-V_{SP}||}{V_{SP}} d\tau\\
            & \text{End-point constraints:}\\
            & (1 - \varepsilon/100)   V(0) \le V(T_F) \le (1 + \varepsilon/100)   V(0)\\
            & \varepsilon_{\pi_\theta} \le  \varepsilon / 100\\
            & \text{Interior-point constraints:}\\
            & a_t = \pi_\theta (F_{in,t}, V_{SP}, V_t, V_{t-1})~\forall t \in \{0, ..., n_T -1\}\\
        \end{split}
    \end{equation}
    
    where \(\tau\) is the continuous time variable, \(V(t)\) is the volume of liquid in the tank, \(F_{in}(t)\) and \(F_{out}(t)\) are the inlet and outlet flows in~\si{m^3.s^{-1}}, respectively, \(a\) is the valve position (1 is open, 0 is closed), and \(f\) is the frequency of the sinusoidal wave disturbance in~\si{s^{-1}}. \(V_{SP}\) is the volume set point, \(\overline{V}\) is the volume of the tank, \(\varepsilon_{\pi_\theta}\) is the integral set point error, \(\varepsilon\) is the maximum allowable error and is equal to 1 \%, \(a_{t}\) is the control action output from the policy applied to the system. 
    
    Constraints are enforced in the formulation to ensure the initial and final states are the same, within 1 \% of error threshold, to achieve cyclic operation which allows for extrapolation of the operation to larger time horizons. The same error threshold, is also used for the maximum setpoint deviation. 

    The goal of the design problem is to determine the maximum deviation for which the controller is able to maintain the desired set-point. While the goal of the OCP is set-point tracking. The OCP is presented in (\ref{eq4}).
    \begin{equation}\label{eq4}
        \begin{split}
            \max_{\pi_\theta} &J_{OCP} = -10  \sum_{t=0}^{n_T}(V_t - V_{SP})^2\\
            &\text{subject to:}\\
            &\dot{V}(\tau) = F_{in}(\tau) - F_{out}(\tau)\\
            & F_{out}(\tau) = \alpha_t   V(\tau)\\
            & F_{in}(\tau) = \hat{F} + F_d   \sin{(\tau / f)}\\
            & f = \frac{1}{2\pi}\\
            & V_{SP} = \hat{F} + F_d \le \overline{V}\\
            & a_t = \pi_\theta (F_{in,t}, V_{SP}, V_t, V_{t-1})~\forall t \in \{0, ..., n_T -1\}\\
        \end{split}
    \end{equation}
    
    %
    %
    where \(J_{OCP}\) is the sum of negative squared setpoint errors, therefore, the maximum reward that is achievable is 0. This objective function is also used for the training to calculate the total reward earned at the end of each episode.
    
\begin{figure}
        \centering
        \includegraphics[width=14cm]{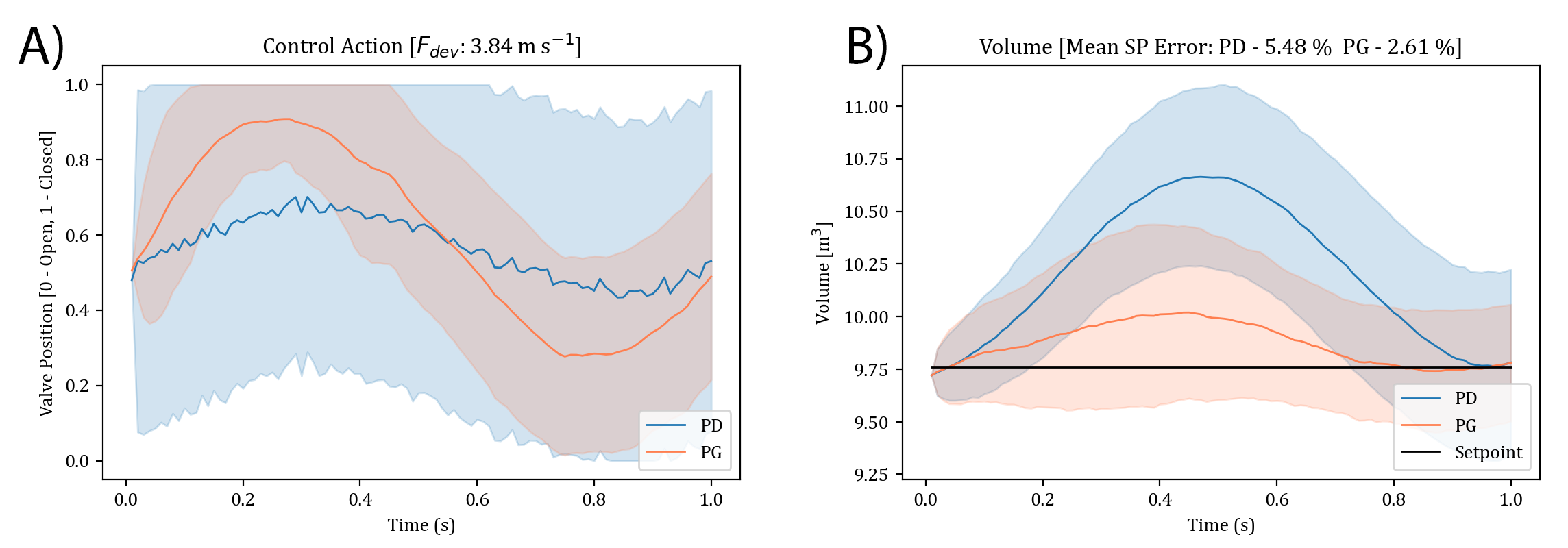}
        \caption{Tank case study control performances, A) control action, B) volume of liquid inside the tank}
        \label{tank}
    \end{figure}

    In this case study, a proportional derivative (PD) controller is used to compare against the approach proposed in this work. The final results showed that the PG controller was able to handle deviation values of up to 3.84~\si{m^3.s^{-1}}. The control performances are plotted in Figure~\ref{tank}.
    
    Figure~\ref{tank} shows the performance of the controllers in the tank case study at the MIDO solution averaged over 1000 simulations with 2 \% measurement noise. It can be observed that the RL controller outperforms the PD controller. The PD control action shows a large variance on its control action indicating very chaotic response. The mean set-point error of the PD controller is two times larger than the RL controller mean error. Furthermore, the RL controller were able to perform better than the mp-MPC used by~\citet{diangelakis2017} ($F_d=2.69 m s^{-1}$) as the RL controller managed to handle a higher maximum deviation even with the presence of measurement noise.
    
    \subsection{CSTR Design}
    \label{cstrdesign}
        The second case study considers the design of a CSTR (continuous stirred tank reactor). The reaction in the CSTR is a first-order endothermic reaction and a heating jacket is equipped. The temperature of the jacket is controlled by the controller. The feed flowrate and concentrations are disturbed by a sinusoidal wave similar to the tank case study. The value of inlet feed flowrate is hidden from the controller. In this case study, there is an option to have a settling tank before the CSTR to normalize the feed mass flowrate to its nominal value. The dynamic system can be found in~\citep{diangelakis2017}. The design problem formulation is shown in (\ref{eqCSTR5}).
         \begin{equation}\label{eqCSTR5}
             \begin{split}
                 \min_{V, m_d, \hat{m}, C_{A_d}, \hat{C_A}, Y_S} &J_{SDC} = Cost_{T}\\
                  &\text{subject to:}\\
                & \dot{C_A} = \frac{m}{\rho V}(C_{A_0}-C_A)- k_0 C_Ae{^{-\frac{E_A}{RT}}}\\
                 &\begin{split}
                     V \rho C_P\dot{T} = mC_P(T_0 - &T) + UA(T_H - T)\\
                     & + V \Delta H_{rxn} k_0 C_Ae^{-\frac{E_A}{RT}}\\
                 \end{split}\\
                 & C_{A_0} = \hat{C_{A_0}} + C_{A_{0_d}} \sin{(2\pi t/100)}\\
                 & m = \hat{m} + m_d   (1-Y_{S,t}) \sin{(2\pi t/100)}\\
                 & T_{H,t} = \pi_\theta (C_{A,t}, C_{A,t-1}, T_t, V, C_{A_0,t}, C_A^{SP})~\forall t\\
                 & \varepsilon_{\pi_\theta}=\int_0^{T_F} C_A - C_A^{SP} d\tau\\
                 & C_A^{SP} = 0\\
                 & \text{Objective function:}\\
                 & Cost_{T} = Cost_{E} + Cost_{O}\\
                 & Cost_{E} = 10((V-750)/\pi) + 1000 + 400 Y_{S,f}\\
                 & \frac{dCost_{O}}{d\tau} = - m (C_{A_0}- C_A) + 4 Y_S\\
                 & \text{Endpoint constraints:}\\
                 & \varepsilon_{\pi_\theta} \le 100,~C_{A_{0_d}} \le \hat{C_{A_0}},~m_d \le \hat{m}\\
                 & \text{Interior point constraints:}~\forall t \in \{0, ..., n_T\}\\
                 & Y_{S,t} - Y_{S,f} \le 0,~Y_{S,t} \in \{0,1\},~Y_{S,f}\in \{0,1\}\\
                 & T_t \le 450\\
             \end{split}
         \end{equation}
        
          where \(C_A\) and \(C_{A_0}\) are the concentrations of A in the reactor and in the feed, respectively, \(m\) is the mass flowrate into and out of the reactor, \(\rho\) is the density of the liquid, \(V\) is the volume of the reactor, \(k_0\) is the pre-exponential factor, \(E_A\) is the activation energy, \(R\) is the ideal gas constant, \(\Delta H_{rxn}\) is the heat of reaction, \(T\) is the temperature inside the reactor, \(T_H\) is the heating utility temperature, \(C_P\) is the heat capacity, \(UA\) is the overall heat transfer coefficient multiplied by the heat transfer area, and \(Y_{S,t}\) is a piece-wise constant binary variable which determines the utilization of the settling tank while \(Y_{S,f}\) is a binary variable which denote the existence of the settling tank. 
         
          The goal of the design problem is to minimize a cost related objective function while satisfying the temperature constraint. While the goal of the OCP is to minimize the concentration of A by manipulating the heating jacket temperature while keeping the CSTR temperature below 450~\si{K}. The OCP is presented in (\ref{eqCSTR6})
        
         \begin{equation}\label{eqCSTR6}
             \begin{split}
                 \max_{\pi_\theta} & J_{OCP} = -\sum_{t=0}^{n_T} (C_{A,t}-C_A^{SP})- 100   \sum_{t=0}^{n_T} \max (0, T_t - 450)\\
                 &  \text{subject to:}\\
                 & \frac{dC_A}{d\tau} = \frac{m}{\rho V}(C_{A_0}-C_A)- k_0 C_Ae{^{-\frac{E_A}{RT}}}\\
                 ~& \frac{dT}{d\tau} = \frac{ mC_P(T_0 - T) + UA(T_H - T) + V \Delta H_{rxn} k_0 C_Ae^{-\frac{E_A}{RT}}}{V \rho C_P}\\
                 & C_{A_0} = \hat{C_{A_0}} + C_{A_{0_d}} \sin{(t/freq)}\\
                 & m = \hat{m} + m_d \sin{(t/freq)}\\
                 & f=100/(2\pi)\\
                 & T_{H,t} = \pi_\theta (C_{A,t}, C_{A,t-1}, T_t, V, C_{A_0,t}, C_A^{SP})~\forall t\\
                 & C_A^{SP} = 0,~T_t \le 450\\
             \end{split}
         \end{equation}
        
         \begin{figure}[H]
             \centering
             \includegraphics[width=10cm]{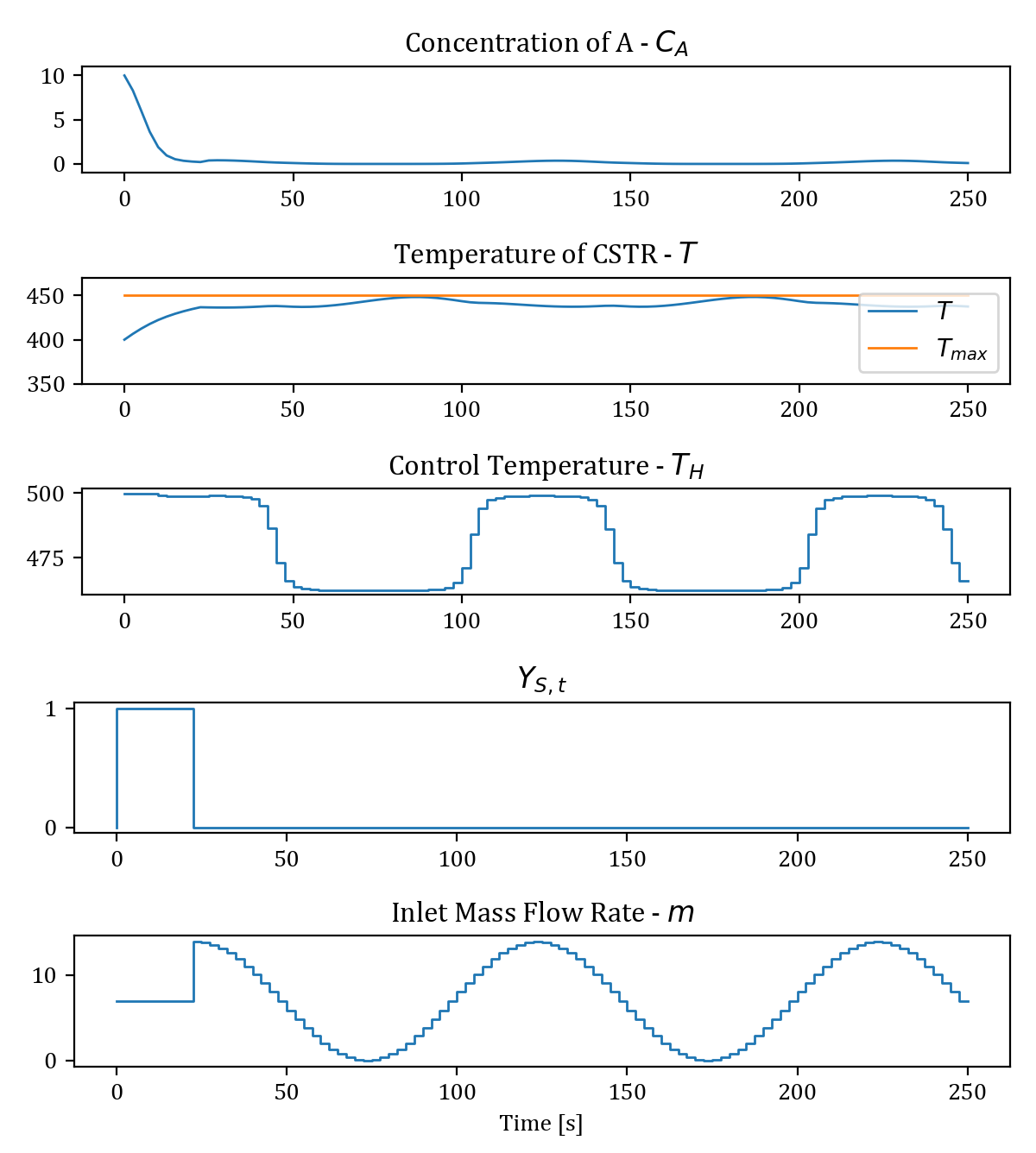}
             \caption{CSTR case study control performance.}
             \label{CSTR}
         \end{figure}
        
         The purpose of the simple PD controller in the tank case study was to speed up the construction of the RL controller. In the CSTR case study however, a PD/PID controller was not considered as it would need complex modifications for it to work with constraints. Tuning the PD/PID to work with the constraints would defeat the purpose of giving a quick head start for the RL controller. The solution has a final cost of 27,539 and the control performance is presented in Figure~\ref{CSTR}. It shows that the settling tank was only needed for a brief period of time at the start of the process. At the start, the concentration of A in the CSTR is higher than the feed concentration, making it hard to satisfy the temperature constraint at initial time, therefore the settling tank was used. Overall, the controller performs very well. It is able to minimize the concentration of A with very good performance (very close to zero almost all of the time) and the temperature constraint is satisfied at all times. This is also true for a wide range of design variables.

\section{Conclusions}
\label{conclusions}
    In this work we proposed the use of RL to address a long standing challenge for simultaneous design and control. This approach is able to address the otherwise intractable nonlinear mixed integer dynamic bilevel optimization problem. Through case studies, we show that the RL-based controller was able to perform well in different design problems enabling the optimization of the process. Policy gradient-based RL controllers can handle measurement noise and stochastic systems naturally and the final form is an explicit function, which can be directly embedded into optimization problems. In future work, we will consider the handling of constraints with high probability such as in~\citep{petsagkourakis2020chance}, and further explore the use of hyperparameter tuning schemes (e.g ~\citet{snoek2012} Bayesian optimization or other expensive black box optimization techniques). We also aim to address harder and larger problems in the near future.


\paragraph{Acknowledgments}
This project has received funding from the EPSRC grant project (EP/R032807/1) is gratefully acknowledged.







\bibliography{references}


\end{document}